\documentclass[showpacs,oneside,twocolumn,prl,amsmath,amssymb,superscriptaddress]{revtex4-1}

\usepackage{cases}
\usepackage{amsmath}
\usepackage{amssymb}
\usepackage{amsfonts}
\usepackage{amssymb}
\usepackage{dcolumn}
\usepackage{bm}
\usepackage{bbm}
\usepackage{graphicx}
\usepackage{xcolor}
\usepackage{array}
\usepackage{subfigure}
\usepackage{hyperref}
\usepackage[normalem]{ulem}

\hypersetup{colorlinks=true,
            breaklinks=true,
            pdfstartview=Fit,
            linkcolor=blue,
            citecolor=blue,
            urlcolor=blue}

\begin{document}


\title{Quantum Simulation of Resonant Transitions for Solving the Eigen-problem of an Effective Water Hamiltonian}

\author{Zhaokai Li}\altaffiliation{These authors contributed equally to this work.}
\author{Xiaomei Liu}\altaffiliation{These authors contributed equally to this work.}%
\affiliation{Hefei National Laboratory for Physical Sciences at the Microscale and Department of Modern Physics, University of Science and Technology of China (USTC), Hefei 230026, China}
\affiliation{CAS Key Laboratory of Microscale Magnetic Resonance, USTC, Hefei 230026, China}
\affiliation{Synergetic Innovation Center of Quantum Information and Quantum Physics, USTC, Hefei 230026, China}

\author{Hefeng Wang}\email{wanghf@mail.xjtu.edu.cn}
\affiliation{Department of Applied Physics, School of Science, Xi'an Jiaotong University and Shaanxi \\ Province Key Laboratory of Quantum Information and Quantum Optoelectronic Devices, Xi'an, 710049, China}

\author{Sahel Ashhab}
\affiliation{Qatar Environment and Energy Research Institute, Hamad Bin Khalifa University, Qatar Foundation, Doha, Qatar}

\author{Jiangyu Cui}
\affiliation{Hefei National Laboratory for Physical Sciences at the Microscale and Department of Modern Physics, University of Science and Technology of China (USTC), Hefei 230026, China}
\affiliation{CAS Key Laboratory of Microscale Magnetic Resonance, USTC, Hefei 230026, China}
\affiliation{Synergetic Innovation Center of Quantum Information and Quantum Physics, USTC, Hefei 230026, China}

\author{Hongwei Chen}
\affiliation{High Magnetic Field Laboratory, Chinese Academy of Sciences, Hefei 230031, China}

\author{Xinhua Peng}\email{xhpeng@ustc.edu.cn}
\author{Jiangfeng Du}\email{djf@ustc.edu.cn}
\affiliation{Hefei National Laboratory for Physical Sciences at the Microscale and Department of Modern Physics, University of Science and Technology of China (USTC), Hefei 230026, China}
\affiliation{CAS Key Laboratory of Microscale Magnetic Resonance, USTC, Hefei 230026, China}
\affiliation{Synergetic Innovation Center of Quantum Information and Quantum Physics, USTC, Hefei 230026, China}


\begin{abstract}
It is difficult to calculate the energy levels and eigenstates of a large physical system on a classical computer because of the exponentially growing size of the Hilbert space. In this work, we experimentally demonstrate a quantum algorithm which could solve this problem via simulated resonant transitions. Using a four-qubit quantum simulator in which two qubits are used as ancillas for control and measurement, we obtain the energy spectrum of a 2-qubit low-energy effective Hamiltonian of the water molecule. The simulated transitions allow the state of the quantum simulator to transform and access large regions of the Hilbert space, including states that have no overlap with the initial state. Furthermore, we make use of this algorithm to efficiently prepare specific eigenstates on the simulator according to the measured eigenenergies. 
\end{abstract}


\maketitle


Calculating molecular energy eigenvalues and eigenstates is an essential task in quantum physics and chemistry. Normally, one has to solve the time-independent Schr\"{o}dinger equation of the system for this purpose. However, numerical simulation of the Schr\"odinger equation on a classical computer is inefficient, because the computational cost grows exponentially with the size of the simulated system~\cite{Levine1999,Thogersen2004}. In quantum chemistry, computational algorithms are usually based on approximate methods that provide rough estimates for chemical properties, and their accuracy varies with the nature of the chemical elements involved. Most of these mathematical algorithms, such as the Variation method~\cite{Szabo1996} and Davidson's algorithm~\cite{Davidson1975}, rely on the diagonalization of the Hamiltonian matrix in an iterative manner with a proper trial state. For large systems, especially with complex interactions, trial states that sufficiently overlap with the desired eigenstates are quite difficult to predict, which generally leads to slow convergence in the iterative computation.

 A quantum computer could efficiently simulate a quantum system by using resources that scale only polynomially with the size of the system~\cite{Feynman1982}. In recent years, quantum simulation of a variety of physical properties has been studied and demonstrated in several experiments~\cite{Georgescu2014,Scalable2016,Bayesian2017,eigensolver2017,DJFH22010,Lanyon2010,lanyon2011universal,kandala2018extending}. For instance, the quantum phase estimation algorithm~(PEA), which can be used to obtain some eigenvalues of a Hamiltonian with exponential speedup over classical algorithms~\cite{Aspuru2005,IPEA2007}, has been implemented to determine the ground-state energy of the Hydrogen molecule~\cite{DJFH22010,Lanyon2010}. The success probability of the PEA depends on the overlap between the trial state and the desired energy eigenstate. The initial guess of the wave function is usually based on either polynomial-scaling classical $ab~initio$ methods~\cite{Veis2010,abinitio2008} or the output of the adiabatic state preparation (ASP) method~\cite{Aspuru2005}. For complicated or excited many-body systems, it is very difficult to guess and efficiently prepare a good trial state for almost any desired eigenstate on a quantum computer.

 Here, we report a nuclear magnetic resonance~(NMR) demonstration of a quantum algorithm for determining the energy levels and eigenstates of a physical Hamiltonian in a given energy range via the simulation of controlled resonant transitions~\cite{WHF2012,berkley2013tunneling,babbush2014adiabatic,WHF2016}. The algorithm does not require preparing an approximation for any particular eigenstate of the simulated Hamiltonian. Instead we start the experiments from a simple reference state and simulated the resonance transitions from this initial state to the eigenstates of the physical Hamiltonian of interest. In the experiment we determine the eigenenergies of the water molecule and then use this information to prepare the corresponding eigenstates on the quantum simulator.

\begin{figure}[h]
\includegraphics[width= 8cm]{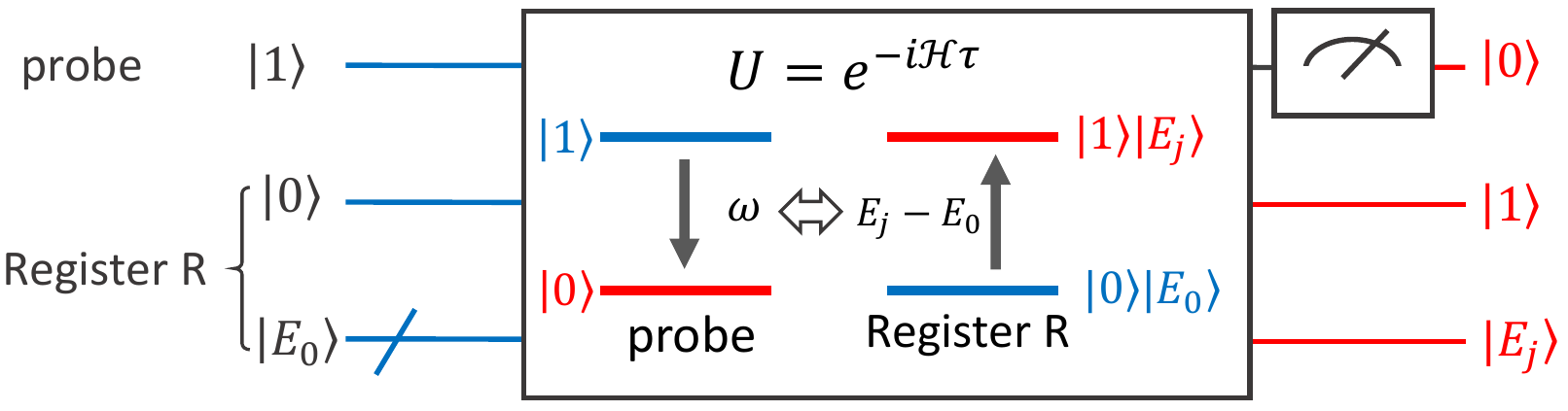}
\caption{(Color online) Schematic diagram of the quantum algorithm. The bottom two lines denote register $R$ consisting of one ancillary qubit and an $n$-qubit system. The evolution governed by the Hamiltonian $\mathcal{H}$ in Eq.~\eqref{wholeH} can drive the register $R$ to resonate with the probe qubit.}
\label{Fig:circuit}
\end{figure}

 The basic idea for applying the algorithm to a physical system is schematically shown in Fig.~\ref{Fig:circuit}. The quantum simulator comprises a probe qubit coupled to a $(n+1)$-qubit quantum register $R$. The register $R$ consists of one ancillary qubit and $n$ system qubits that encode the simulated system. The probe qubit exhibits a dynamical response only when it is resonant with a transition between a pair of energy eigenstates of the register $R$ with the same energy separation. The non-invasive measurement via the probe qubit is crucial for the use of the algorithm for eigenstate preparation and for performing a sequence of iterations through which the state of the simulator can move far away in the Hilbert space from the initial state.
 
 The Hamiltonian of the simulator has the form
\begin{equation}
    \mathcal{H} = - \frac{\omega}{2} {\sigma_z} \otimes \mathbf{I}_2^{ \otimes \left( {n + 1} \right)} + {\mathbf{I}_2} \otimes {\mathcal{H}_R} + c{\sigma _x} \otimes A \label{wholeH},
\end{equation}
where ${\mathbf{I}_2}$ is the two-dimensional identity operator, and $\sigma_{x,y,z}$ are Pauli matrices. The first term in Eq.~\eqref{wholeH} denotes the Hamiltonian of the probe qubit which has an energy gap $\omega$~($\hbar=1$). The second term represents the Hamiltonian of register $R$ and is given by
\begin{equation}\label{HamiltonianR}
 {\mathcal{H}_R} = \left\vert 0 \right\rangle \left\langle 0 \right\vert \otimes \mathcal{H}_{E_0} + \left\vert 1 \right\rangle \left\langle 1 \right\vert \otimes {\mathcal{H}_S}.
\end{equation}
 Here $\mathcal{H}_{E_0}$ is defined as $\mathcal{H}_{E_0}=E_0\left\vert E_0 \right\rangle \left\langle E_0 \right\vert$ with $E_0$ serving as an energy reference point and the eigenstate $\left\vert E_0 \right\rangle$ being the input state. This state does not need to have substantial overlap with any particular target eigenstate. The transitions that take place during the implementation of the algorithm can transform the state of the simulator to states that have no overlap with the initial state. It should also be noted here that if one has a good guess state, e.g. one obtained using coupled-cluster or tensor-network methods, it can be used as an initial state to enhance the efficiency of the algorithm. ${\mathcal{H}_S}$ is the Hamiltonian of the simulated system whose Hilbert space size is $N=2^n$ and $\left\vert {{E_j}} \right\rangle $ denotes the $j$th eigenstate with eigenenergy $E_j$, $j=1,...,N$.

The third term in Eq.~\eqref{wholeH} describes the interaction between the probe qubit and the register $R$. The coefficient $c$ is the coupling strength, which should be set to a value that is much smaller than $\omega$ to suppress off-resonant transitions. The operator $A$ is defined as $A=\sigma_x \otimes B$ where $B$ is the operator that will drive the transformation in the system qubits.

In the initial state, we prepare the probe qubit in its excited state $\left\vert 1 \right\rangle$ and the $(n+1)$-qubit register $R$ in the reference state $\left\vert 0E_0 \right\rangle $. Then, the quantum simulator evolves under the Hamiltonian $\mathcal{H}$ with a given $\omega$ for time $\tau$. When the energy gap between $E_0$ and $E_j$ equals the gap of the probe qubit, i.e., $\omega=E_j-E_0$, the simulator is on resonance, which leads to oscillations where the probe qubit is de-excited from $\left\vert {1} \right\rangle$ to $\left\vert {0} \right\rangle$ while register $R$ is excited from $\left\vert 0E_0 \right\rangle $ to $\left\vert {{1E_j}} \right\rangle $. The transition probability $P$ is given by
\begin{equation}
 \label{Pdecay}
  {P_j(\omega)} = \frac{(2c|d_j|)^2 }{\Omega_{0j}^2}{\sin ^2}\left( {\frac{{{\Omega _{0j}}\tau }}{2}} \right),j = 1,2,...,N,
\end{equation}
where $d_{j}=  {\left\langle {{1E_j}} \right \vert A\left\vert {{0E_0}} \right\rangle } = {\left\langle {{E_j}} \right\vert B\left\vert {{E_0}} \right\rangle } $ and $\Omega_{0j}={\sqrt {(2c|d_j|)^2 + {{\left( {{E_j} - {E_0} - {\omega}} \right)}^2}}}$. 
After the time evolution we measure the probability $P(\omega)$ of the probe qubit to be in state $\left\vert 0\right\rangle$. Absorption peaks appear in $P(\omega)$ at $\omega$ values that are close to $E_j-E_0$. Hence, the observation of a peak at frequency $\omega$ can be seen as the measurement of one energy eigenvalue of $\mathcal{H}_S$ with the value ${E_j}={E_0}+{\omega}$.  It should be noted that the performance of the algorithm can be optimized by choosing appropriate values for the coupling strength $c$, transition operator $B$ and evolution time $\tau$~\cite{WHF2016}. For example, one could first run the algorithm with a relatively large value of $c$, which would lead to broad and easily detectable resonances. One could then rerun the algorithm with smaller values of $c$ and focus on the regions where the broad resonances were observed to determine the locations of the resonance peak centers with higher accuracy. It should also be noted that, if one chooses a relatively large value of $c$, one must be careful to choose other parameters, e.g. the Trotter time step, accordingly, e.g. to keep errors from growing beyond the desired measurement accuracy.

 By setting $\omega$ at the resonant frequency $\omega_j=E_j-E_0$ and $c\ll\left\vert{E_j} -{E_i}\right\vert$, which suppresses off-resonant transitions, the final state of the quantum simulator is approximately of the form:
\begin{equation}\label{finalstate}
\begin{aligned}
 \left\vert \Psi \right\rangle =&e^{i\phi_0}\sqrt{1-P_j(\omega_j)}\left\vert  1 \right\rangle \left\vert 0 \right\rangle\left\vert {E_0} \right\rangle \\
 &+ e^{i\phi_1}\sqrt{P_j(\omega_j)}\left\vert 0 \right\rangle\left\vert 1 \right\rangle \left\vert {E_j} \right\rangle,
 \end{aligned}
 \end{equation}
where the phases $\phi_0$ and $\phi_1$ do not lead to any observable effects in the algorithm. In this situation the transition probability $ {P_j(\omega_j)} = {\sin ^2}\left({{c|d_j|\tau }}\right) $, and hence resonant Rabi oscillations in the time domain could be observed if one varies the time $\tau$ with fixed $\omega$. By choosing the appropriate $\tau$ that leads to the maximal $P$, which is 1 on resonance, the desired eigenstate $| E_j\rangle$ can be prepared deterministically.

In the experiment, we implement the algorithm for the water molecule as an example. Considering the electronic structure of the water molecule and multi-reference-configuration interaction (MRCI) calculations~\cite{Szabo1996,abinitio2008,WHF2012}, we encode the low-energy Hamiltonian $\mathcal{H}_S$ of the water molecule in a four-dimensional Hilbert space. Details of the Hamiltonian matrix calculation are described in the Supplementary Material~\cite{Supplement}. In this article, energies and time are recorded in units of Hartree and Hartree${}^{ - 1}$, respectively. We set $\left\vert E_0\right\rangle =\left\vert 00\right\rangle $. Hence, the Hamiltonian of the quantum simulator in Eq.~\eqref{wholeH} has the form
\begin{equation}
\label{H in circuit}
\begin{aligned}
  \mathcal{H}= - &\frac{\omega}{2} {\sigma _z} \otimes \mathbf{I}_2^{ \otimes 3}+ c{\sigma _x} \otimes A\\
  &+ {\mathbf{I}_2} \otimes \left(\left\vert  0 \right\rangle \left\langle 0 \right\vert\otimes E_0\left\vert 00 \right\rangle \left\langle 00 \right\vert + \left\vert 1 \right\rangle \left\langle 1 \right\vert \otimes {\mathcal{H}_S}\right),
\end{aligned}
\end{equation}
in which the three terms represent the Hamiltonian of the probe, the interaction Hamiltonian and the Hamiltonian of register $R$, respectively. Here, $A= {\sigma _x} \otimes B= {\sigma _x} \otimes H_d \otimes H_d$ acts as the transition operator with the Hadamard operator $H_d$, which can transform register $R$ from the reference state $\left\vert 0 \right\rangle \otimes \left\vert 00 \right\rangle $ to the state ${\left\vert 1\right\rangle }\otimes\frac{1}{2^2}\sum\nolimits_{k = 0}^{3} {\left\vert k \right\rangle } $ covering all the computational basis states in the subspace of $\left\vert 1 \right\rangle \!\! \left\langle 1 \right\vert \! \otimes \!\mathcal{H}_S$.

\begin{figure}
\includegraphics[width= 8cm]{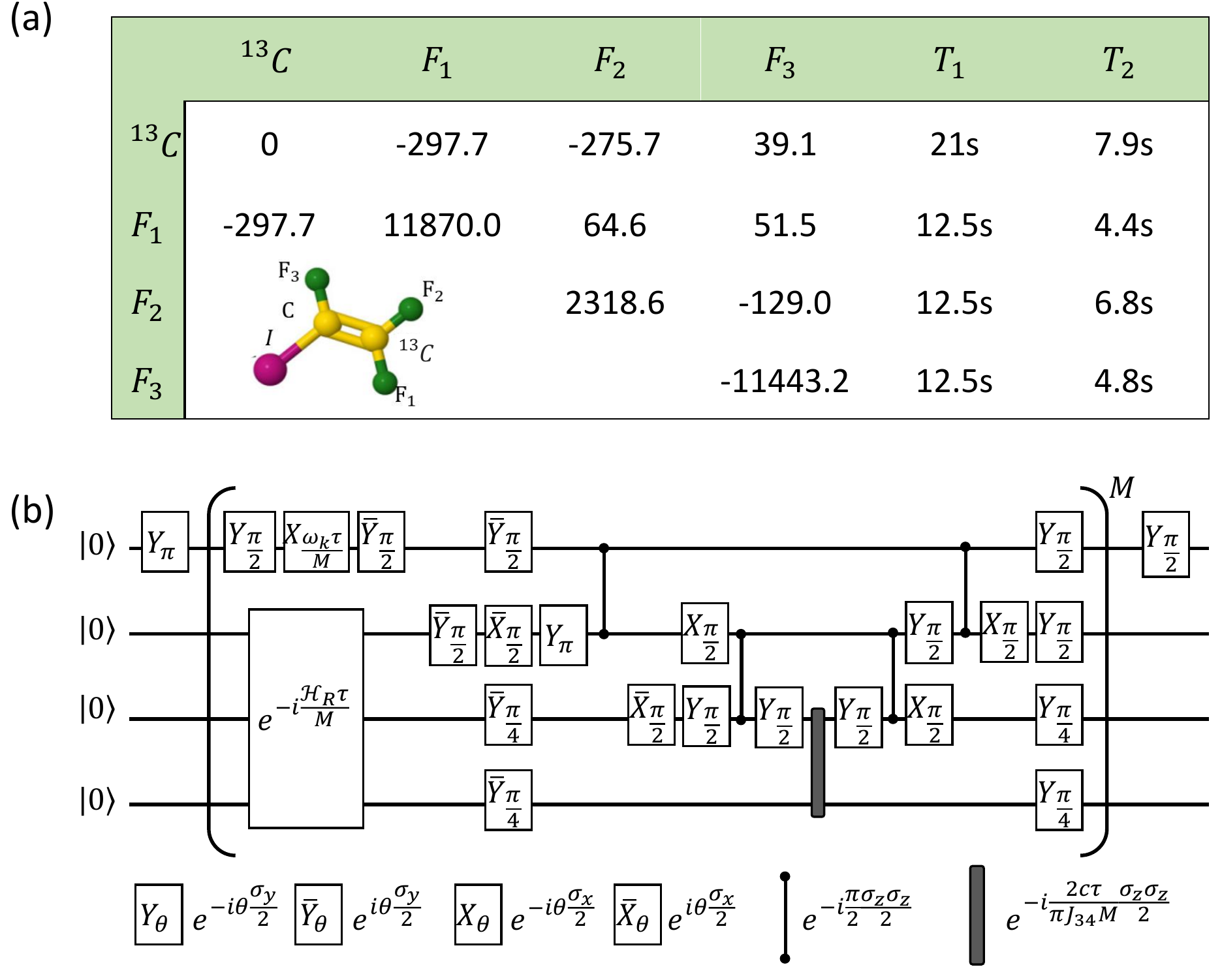}
\caption{(Color online)(a) Molecular properties of ${}^{13}\text{C}$-iodotrifluoroethylene ($\rm{C}_2\rm{F}_3\rm{I}$). The chemical shifts ${\nu_j}$ and the scalar coupling constants ${J_{jk}}$~(in units of Hz) are given on and above the diagonal of the table, respectively. The spin-lattice relaxation time $T_1$ and spin-spin relaxation time $T_2$ are in the last two columns. (b) Quantum circuit to implement the algorithm. The logic gate sequence between parentheses performs the operator ${e^{ - i\mathcal{H}\tau/M}} $, where $M$ denotes the number of time steps and ${\mathcal{H}}$ is the Hamiltonian of the simulator.}
\label{Fig:sample}
\end{figure}

 The experiments were carried out on a Bruker AV-400 NMR spectrometer at room temperature. We used ${}^{13}\rm C$-iodotrifluoroethylene ($\rm{C}_2\rm{F}_3\rm{I}$) dissolved in $d$-chloroform as the quantum simulator, with a ${}^{13}\rm{C}$ nuclear spin acting as the probe qubit, a ${}^{19}\rm{F}_1$ nuclear spin as the ancillary qubit and ${}^{19}\rm{F}_2$ and ${}^{19}\rm{F}_3$ nuclear spins as qubits for the simulated system. The natural Hamiltonian of this four-nuclear-spin liquid-state NMR system in the doubly rotating frame is
\begin{equation}
\label{sampleH}
 {H_{\rm{NMR}}} = \sum\limits_{j = 1}^4 {{\pi\nu _j}\sigma _z^j}  + \sum\limits_{1 \leqslant j < k \leqslant 4} {\frac{\pi }{2}{J_{jk}}\sigma _z^j\sigma _z^k},
\end{equation}
 where $\nu _j$ is the chemical shift of spin $j$, and $J_{jk}$ represents the coupling constant between spin $j$ and spin $k$. The molecular structure and Hamiltonian parameters of this sample are shown in Fig.~\ref{Fig:sample}(a). The procedures of the experimental implementation are described as follows.

 (i) \emph{Preparation of the initial state}. Starting from the thermal equilibrium state of the four-qubit quantum simulator, we employ the line-selective method to prepare the pseudopure state~(PPS)~\cite{Gershenfeld1997,PXH2001} $\rho = \left( {1 - \varepsilon /16} \right){{\mathbf{ I}}_{2}^{\otimes4}} + \varepsilon \left\vert {0000} \right\rangle \left\langle {0000} \right\vert$, in which $\varepsilon  \approx {10^{ - 5}}$ represents the polarization. Then, by applying a $\pi$ radio-frequency pulse along the $Y$ axis to the probe qubit, we prepare the simulator in the initial state $ \left\vert \Psi \right\rangle_0 = \left\vert {1000} \right\rangle$ where the probe qubit is in the excited state $\left\vert 1 \right\rangle $ and register $R$ remains in $\left\vert {000} \right\rangle $ serving as the reference state.

(ii) \emph{Implementation of the evolution operator}. We set the reference energy at $E_0=-84.20$, the scanning frequency range of the probe qubit as $\left[ {{\omega _{\min }} , {\omega _{\max }}} \right]= \left[ {0,2} \right]$, and the parameters in the Hamiltonian $\mathcal{H}$ at $c=0.006$ and $\tau=1000$. By dividing the frequency range into $100$ intervals of width $\Delta \omega  = 0.02$, we form a scanning data set for the probe qubit with ${\omega _k} = {\omega _{\min }} + {k}\Delta \omega $, $k=1,2,...,100$. The quantum circuit to implement the evolution operator $U=e^{-i\mathcal{H}\tau}$ using the Trotter formula \cite{reiher2017elucidating, casanova2012quantum, poulin2015trotter} is shown in Fig.~\ref{Fig:sample}(b)~\cite{Oritz2001,Tseng1999}, keeping in mind that some studies have suggested that alternative methods such as sparse-matrix \cite{berry2007efficient} and Taylor-series methods \cite{berry2015simulating} might be more efficient for larger systems. In the experiment, the evolution operator for each $k$ is implemented using a shaped pulse which has a duration of 20 ms and is optimized by using the gradient ascent pulse engineering technique~\cite{Navin2005}. All the shaped pulses have theoretical fidelities above 99.5\% and are robust against the inhomogeneity of radio-frequency pulses.

\begin{figure}
\includegraphics[width= 8cm]{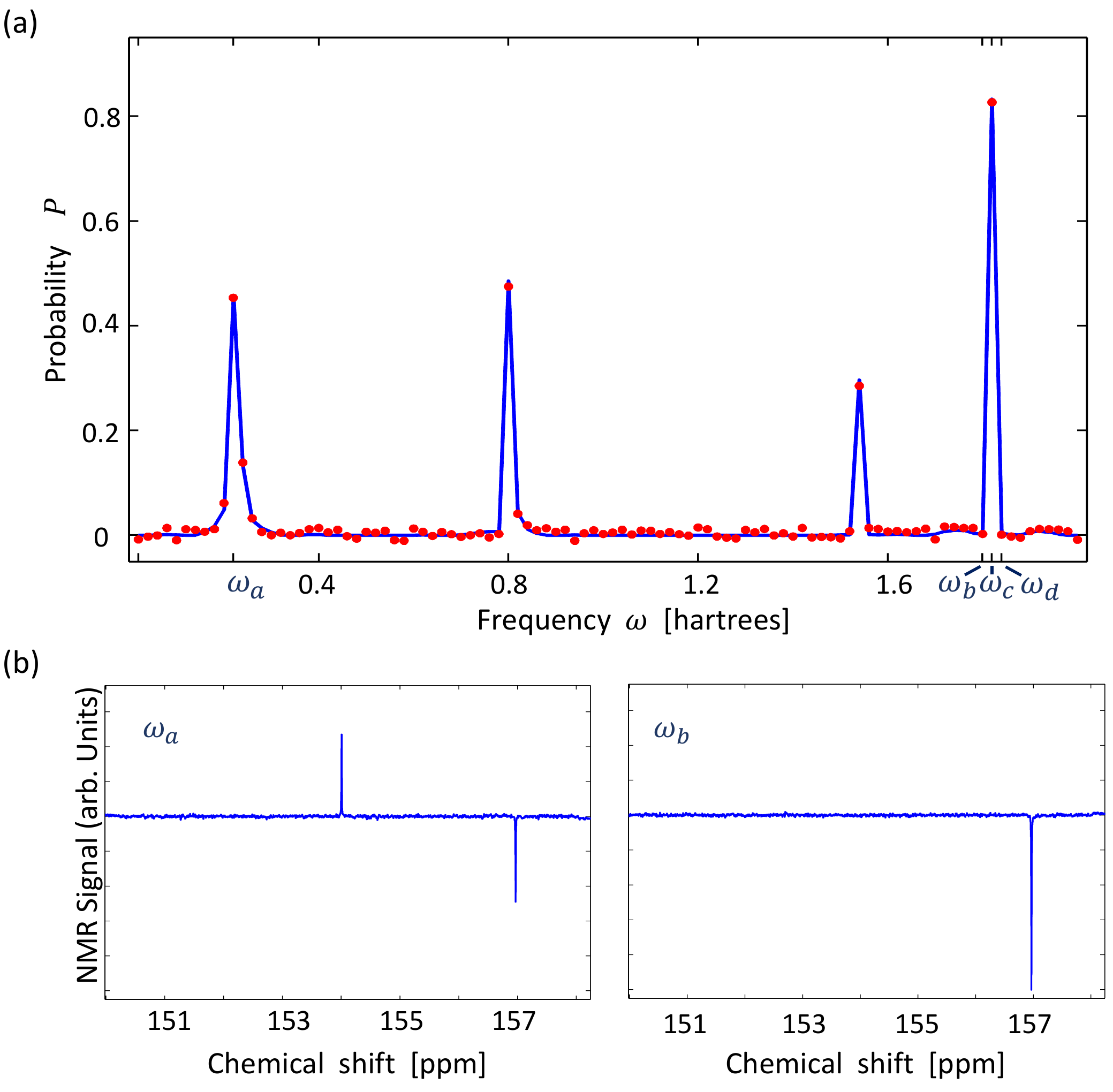}
\caption{(Color online)(a) Energy spectrum of the water molecule obtained using the quantum algorithm. The blue lines and red dots represent the theoretical and experimental probability $P$ of the probe qubit in the state $\left\vert 0 \right\rangle $ as a function of frequency $\omega$. Here, $\omega_{a,b,c,d}=0.22$, $1.80$, $1.82$, $1.84$. (b) Experimental NMR signal spectrum of $^{13}C$ in the final state at $\omega_a=0.22$ and $\omega_b=1.80$~(both indicated in~(a)).}
\label{Fig:energyspectrum}
\end{figure}

 (iii) \emph{Measurement of the energy spectrum}. The final state of the simulator after evolution is approximately given by Eq.~\eqref{finalstate}. When we trace out the subspace of the register $R$, the final density matrix of the probe qubit is reduced to ${\rho'} = \left( {\begin{array}{*{20}{c}}
  P_j(\omega)&0 \\
  0&{1 - P_j(\omega)}
\end{array}} \right)$. By applying the readout pulse $Y_{\frac{\pi}{2}}$ to the probe qubit, the ensemble average probability $P_j(\omega)$ of the probe qubit in state $\left\vert 0\right\rangle$ is measured from the NMR signal spectrum~(Fig.~\ref{Fig:energyspectrum}(b)).

 By scanning the frequency $\omega$ of the probe qubit through the discrete set $\omega_k$, we obtain the energy spectrum of the water molecule shown in Fig.~\ref{Fig:energyspectrum}(a). Figure \ref{Fig:energyspectrum}(b) shows the NMR signal spectrum in two different cases. For the resonant frequency setting $\omega_a=0.22$~(the first peak in Fig.~\ref{Fig:energyspectrum}(a)), the probe qubit is de-excited from state $\left\vert 1 \right\rangle $ to state $\left\vert 0 \right\rangle $ with probability $0.4531$, which indicates that one of the eigenstates of the system has the eigenvalue $E_1=E_0+\omega_a=-83.98$. In contrast, the spectrum at the non-resonant frequency $\omega_b=1.80$ shows that the quantum simulator remains in the state $\left\vert 1000 \right\rangle $ with the transition probability near zero, indicating that there is no resonance at this point. In total, there are four peaks observed in the energy spectrum, and thus four energy eigenvalues of the water molecule are measured as $-83.98$, $-83.40$, $-82.66$, $-82.38$, which are in good agreement with theoretical expectations: $-83.9731$, $-83.4010$, $-82.6604$, $-82.3763$. The difference between the theoretical and experimental values, which is below $0.01$ for all four energies, is limited by $\Delta \omega$. More accurate results can be obtained by focusing on the resonance peaks obtained in the first run of the algorithm and repeating the algorithm with smaller values of $c$ and $\Delta\omega$.

\begin{figure}
\includegraphics[width= 8cm]{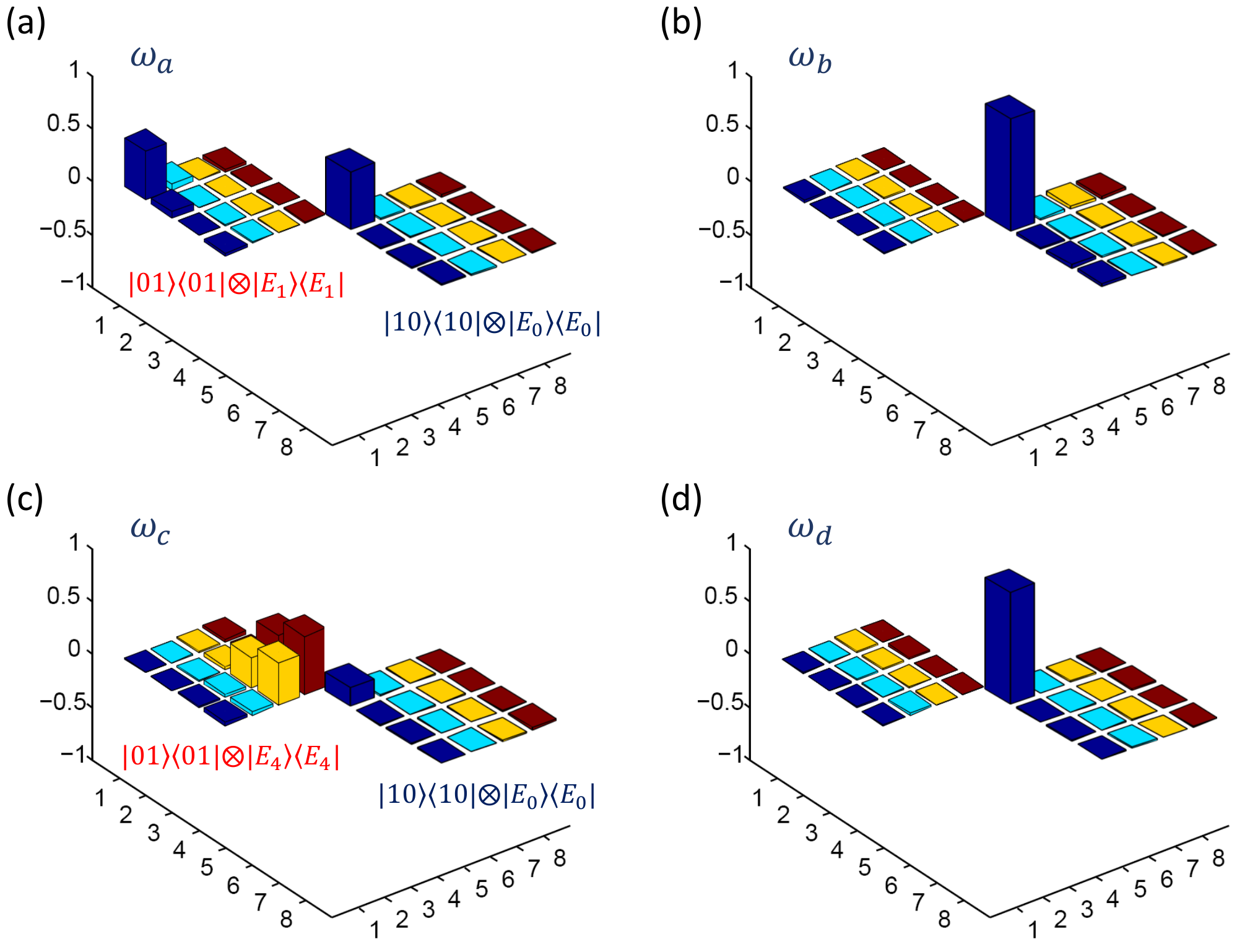}
\caption{(Color online) Experimental state tomography of the quantum simulator: (a)-(d) respectively show the real parts of experimental reconstructed density matrices at frequencies $\omega_{a,b,c,d}$ indicated in Fig.~\ref{Fig:energyspectrum}(a). The rows and columns labeled 1-4 represent computational basis states from $\left\vert {0100} \right\rangle$ to $\left\vert {0111} \right\rangle$, and those labeled 5-8 represent basis states from $\left\vert {1000} \right\rangle$ to $\left\vert {1011} \right\rangle$. The subspaces $\left\vert 00\right\rangle_{12}$ and $\left\vert 11\right\rangle_{12}$ appear with negligible probabilities and are not shown. All imaginary parts of the density matrices are also negligible.}
\label{Fig:statetomo}
\end{figure}

(iv) \emph{Preparation of eigenstates}. As discussed above, once an energy $E_j$ has been identified, one can run the algorithm with appropriately set parameters to deterministically prepare the state $|E_j\rangle$. Alternatively, one can also resort to a heralded state preparation approach, keeping only those instances in which the probe and ancillary qubits are in the state $|01\rangle_{12}$. The heralded approach can be useful for discarding some of the unsuccessful instances in the presence of decoherence and imperfections in the implementation. Here, we have implemented this protocol, as we demonstrate using a tomography procedure ~\cite{Supplement,Lee2002}.
Figure~\ref{Fig:statetomo} shows that the experimentally reconstructed density matrix of the final state for different values of $\omega$. At the resonant frequency $\omega_a=E_1-E_0=0.22$, the system makes a transition from the reference state $\left\vert E_0\right\rangle$ in the subspace of ${\left\vert {10} \right\rangle}_{12}$ to the ground state $\left\vert E_1\right\rangle$ in the subspace of ${\left\vert {01} \right\rangle}_{12}$ with the probability $P_1(\omega_a)=0.4531$. After renormalization, the density matrix in the subspace of ${\left\vert {01} \right\rangle}_{12}$ has a closeness \cite{correlation_note} of $99.32\%$ with the ground state of the water molecule, indicating that the system is faithfully prepared in the ground state when the transition occurs. The excited state $\left\vert E_4\right\rangle$ is prepared by setting the frequency at $\omega_c=E_4-E_0=1.82$, with a closeness of $99.34\%$ and a success probability of $0.8264$. For comparison, when we set the frequency outside the resonance regions, such as $\omega_b=1.80$ or $\omega_d=1.84$, the simulated system remains in its initial state $\left\vert 10E_0\right\rangle$~(Fig.\ref{Fig:statetomo}(b,d)). 

The deviations of the experimental final states from the expected eigenstates result mainly from two factors: (1) the fluctuations of the strength of the NMR signal that are around $1\%$, (2) the imperfection of GRAPE pulses with theoretical infidelities up to $\sim$ 0.5\% which leads to a fluctuation of the spectra of around $2\%$. The high fidelities in the tomography data confirm the feasibility of the quantum algorithm for preparing eigenstates.

As we have mentioned above, the input state $\left\vert E_0\right\rangle$ of our algorithm does not need to be a good approximation to the eigenstate of interest. Moreover, since the eigenstates can be prepared deterministically and non-destructively in our algorithm, one can use the eigenstates that are prepared in the first run of the algorithm as input states for further transitions. By allowing a sequence of multiple transitions, the quantum simulator can explore a large portion of the Hilbert space, including states that have little or no overlap with the initial state. In contrast, when using the PEA for example, the algorithm only projects the initial state onto different energy eigenstates that have substantial overlap with the initial state. The transition operator $B$ is one of the inputs that can be varied in the algorithm. While we have chosen a simple operator in this experiment, the operator could be designed based on any information that we might have about the target state, or it could be varied in different runs to reduce the probability of missing a transition with a very small matrix element. 

Considering the scaling of the algorithm, two factors should be considered: (1) the number of experiments that need to be performed to obtain an energy spectrum in a given scanned energy range. (2) the time needed to run the circuit in each experiment. The first factor, i.e. the number of experiments, scales as $O(1/\epsilon^2)$ where $\epsilon$ denotes the desired uncertainty of the eigenvalue $E_j$. The second factor is determined by the computational cost of implementing the unitary operation $e^{-i H \tau}$, which obeys the rules of implementing the time evolution operator of a realistic system, with a standard additional overhead for the ancillary and probe qubits. Although the exact scaling laws of simulating a realistic system are not fully known and can vary depending on the details of the implementation \cite{kivlichan2018quantum,heyl2018quantum}, one can safely say that the computational resources required scale polynomially with system size, as opposed to the exponential scaling expected for classical computers. The resources required also scale polynomially with the desired accuracy $\epsilon$ of the eigenvalues \cite{Supplement}.

 To conclude, we have experimentally demonstrated a quantum algorithm for calculating the eigenenergies and eigenstates of a known Hamiltonian without having a good initial approximation for any of its eigenstates. As an example, we successfully obtained the lowest four energy eigenvalues of the water molecule by simulating resonant transitions between the system and a probe qubit, and prepared the corresponding eigenstates with high fidelity on a four-qubit NMR quantum simulator. Our quantum algorithm provides an additional option in the set of quantum algorithms for simulating eigenvalue problems and preparing eigenstates of large, complicated Hamiltonians.

\begin{acknowledgments}
This work is supported by National Key R\&D Program of China (Grant No.\ 2018YFA0306600 and 2017YFA0305000), the National Natural Science Foundation of China (Grants  No.\ 11575173, No.\ 11425523, No.\ 11275145, No.\ 11661161018, and No.\ U1632157), the Natural Science Fundamental Research Program of Shaanxi Province of China (Grant No.\ 2018JM1015), and Anhui Initiative in Quantum Information Technologies (Grant No. AHY050000). Z.\ Li thanks the China Scholarship Council for the support.
\end{acknowledgments}

\bibliography{referenceH2O}

\begin{thebibliography}{37}
\bibitem [{cor()}]{correlation_note}%
  \BibitemOpen
  \href@noop {} {}\bibinfo {note} {The closeness between two density matrices
  ${\sigma}$ and ${\rho}$ is defined as the attenuated correlation $C =
  \rm{Tr}\left( {{\sigma}{\rho}} \right)/\sqrt {\rm{Tr}\left( {\sigma^2}
  \right)\rm{Tr}\left( {\rho^2} \right)} $}\BibitemShut {NoStop}%
\end{thebibliography}
\bibliographystyle{apsrev4-1}

\section{Supplementary Material}

\maketitle
\section{The Hamiltonian of the Water molecule}\label{HamiltonianH2O}

 The electronic Hamiltonian of the water molecule in the form of second quantization~\cite{Levine1999} is $H_S = \sum\limits_{p,q} \left\langle p \right|\left( {T + {V_N}} \right)\left| q \right\rangle a_p^ + {a_q}  - \frac{1}{2}\sum\limits_{p,q,r,s} \left\langle p \right|\left\langle q \right|{V_e}\left| r \right\rangle \left| s \right\rangle a_p^{+}a_q^{+}{a_r} {a_s}$ where $|p\rangle$, $|q\rangle$, $|r\rangle$, $|s\rangle$ denote the one-particle states, ${a_p^ + }$ is its fermionic creation operator, and operators $T$, $V_N$, and $V_e$ are the one-particle kinetic energy operator, nuclear attraction operator and two-particle electron repulsion operator, respectively. The Hartree-Fock wave function for the electronic structure of the water molecule is ${(1{a_1})^2}{(2{a_1})^2}{(1{b_2})^2}{(3{a_1})^2}{(1{b_1})^2}$. The number of qubits needed to represent the water molecule on a quantum simulator can be optimized in terms of the state-mapping technique presented in Ref.\cite{abinitio2008}. Considering the $^1A_1$ symmetries and using the STO-$3$G basis set and freezing the first two $a_1$ orbitals, the $b_1$ orbital and the first $b_2$ orbital, we construct a model space that includes the $3a_1$, $4a_1$ and $2b_2$ orbitals for the low-energy Hamiltonian $\mathcal{H}_S$ of water molecule, and perform configuration interaction~(CI) calculations~\cite{Szabo1996}. Accordingly, the multi-reference-configuration interaction (MRCI) space is composed of four configuration state functions which require two qubits to represent the state of the water molecule, and the Hamiltonian matrix is given by~\cite{Davidson1975}:
  \begin{equation}
  \label{systemH}
{\mathcal{H}_S} = \left( {\begin{array}{*{20}{c}}
  { - 83.9566}&{ - 0.0820}&{0.0458}&{0.0594} \\
  { - 0.0820}&{ - 83.4080}&{0.0110}&{0.0767} \\
  {0.0458}&{0.0110}&{ - 82.5661}&{0.1323} \\
  {0.0594}&{0.0767}&{0.1323}&{ - 82.4800}
\end{array}} \right)
  \end{equation}
whose eigenvalues are $-83.9731$, $-83.4010$, $-82.6604$ and $-82.3763$. In this paper, energies and time are recorded in units of Hartree and Hartree${}^{ - 1}$, respectively.

In the general case, one can always write down the Hamiltonian of a chemical system in the second quantization formalism and map it to operators that act on qubits via the Jordan-Wigner transformation. The Hamiltonian contains only a polynomial number of local interaction terms,  which can be transformed to operators that only act on a few qubits, and can be implemented efficiently. Thus, one can safely say that the computational resources required for simulating a chemical system on a quantum computer scale polynomially with system size, as opposed to the exponential scaling expected for classical computers. 

 \section{The tomography procedure}
We implement a tomography procedure to read the final state in the subspaces of ${\left| {01} \right\rangle}_{12}$ and ${\left| {10} \right\rangle}_{12}$. Since the natural abundance of this NMR sample with just one $^{13}\text{C}$ carbon is about $1\%$, we read out all qubits via the $^{13}\rm{C}$ channel to distinguish these molecules against the large background. This partial tomography is achieved by applying eight combined readout pulses: $Y_{\frac{\pi }{2}}\text{III}$, $Y_{\frac{\pi }{2}}\text{III}\ast \text{SWAP}_{12}$, $Y_{\frac{\pi }{2}}\text{III}\ast \text{SWAP}_{13}$, $Y_{\frac{\pi }{2}}\text{III}\ast \text{SWAP}_{14}$ for measuring all diagonal elements, $\text{SWAP}_{13}$, $\text{SWAP}_{14}$ for detecting single-quantum coherences, and $\text{III}Y_{\frac{\pi }{2}}\ast \text{SWAP}_{13}$, $\text{III}X_{\frac{\pi }{2}}\ast \text{SWAP}_{13}$ for measuring zero- and double-quantum coherences, in which $\text{I}$ is the unity operator; $X_{\frac{\pi }{2}}$ and $Y_{\frac{\pi }{2}}$ represent the $\pi/2$ rotations along the $x$ and $y$ axis; $\text{SWAP}_{ij}$ denotes the swap operation between the $i$th and $j$th qubits.

\section{The Scalability of the algorithm}

In our algorithm, we note that the state of the system register changes as a result of the resonant transitions. As a result, by using multiple transitions the simulator can explore a large portion of the Hilbert space, including states that have little or no overlap with any guess state. In contrast, when using the PEA for example, the algorithm is repeated $N$ times, giving the energies of states that have $N|\langle E_j|E_{guess}\rangle|^2>1$. The algorithm only projects the initial state onto different energy eigenstates that have substantial overlap with the initial state.  Thus, the PEA algorithm is sensitive to the overlap between the input state and the measured eigenstates as are other similar algorithms. The algorithm implemented in our work has the advantage that even if the direct transition from the initial input state to a given eigenstate is very weak, one can use eigenstates that are prepared using the algorithm as input states for further transitions. The transformation of the initial state as a result of the resonant transitions is therefore an important advantage of our algorithm.

Considering the scaling of the algorithm, as it essentially performs quantum simulations of the dynamics, it follows the same scaling laws (as a function of system size) as quantum simulation. It therefore gives an exponential speedup over classical algorithms when dealing with large, complicated Hamiltonians. Two factors should be considered in relation to the efficiency of the algorithm: (1) the number of experiments that need to be performed to obtain an energy spectrum in a given scanned energy range. (2) the time needed to run the circuit in each experiment. The total number of experiments is equal to $KL$, while $K$ represents the number of experiments required to have a statistical estimation of the transition probability $P_j(\omega)$ for each frequency $\omega$, and $L$ represents the number of frequency points in the spectrum. If one wishes to determine the eigenvalue with uncertainty $\epsilon$, the total number required will scales as $O(1/\epsilon^2)$. We note that one does not need to cover the entire frequency domain with the same intervals of frequency points in the spectrum, i.e. one can zoom in on the resonances and keep a lower density of frequency points elsewhere. Hence, depending on the problem, one might be able to obtain better scaling than $O(1/\epsilon^2)$. 
 
 The second factor, i.e. the time needed to run the algorithm, is determined by the computational cost of implementing the unitary operation $e^{-i H \tau}$. In the experiment, the system-probe coupling strength is chosen such that $c\left\vert d_j\right\vert\sim\epsilon$. The evolution would then need to be performed for a time $ \pi/(2c\left\vert d_j\right\vert)$, and hence the time will scale as $\tau\sim 1/(2\epsilon)$. The implementation of $e^{-i H \tau}$ obeys the rules of implementing the time evolution operator of the quantum system of interest, i.e. $e^{-i H_s \tau}$, with a standard additional overhead for the control and probe qubits. Since $H_S$ is a physical Hamiltonian, it typically includes only one and two-body interactions and could be represented by $O(N^2)$ second-quantized terms, where $N$ is the number of spin orbitals \cite{babbush2018low}. Although the exact scaling laws of the quantum simulation are not fully known and can vary depending on the details of the implementation \cite{kivlichan2018quantum,heyl2018quantum}, one can safely say that the computational resources required for simulating a physical system scale polynomially with system size, as opposed to the exponential scaling expected for classical computers. 

Furthermore, while we have used an NMR setup, the algorithm is based on the circuit model of quantum computing. It can therefore be implemented on other platforms that support the circuit model. For example, there are now superconducting devices with 10-100 qubits that can be used to implement the algorithm for larger molecules. In those devices it is the coherence time rather than the number of qubits that limits the size of the problem that can be addressed. Judging from recent experiments, implementing our algorithm with 5-10 system qubits seems to be a reasonable estimate for the system size that can be tackled on present-day quantum computers."

\end{document}